          [2001/04/25 1.1 (PWD)]
\documentclass[a4paper,twocolumn]{esapub} % European paper
\usepackage{natbib,graphicx}

\title{The nature of the faint sub-mJy radio population}
\author{Nick Seymour}
\affil{Institut d'Astrophysique de Paris, 98bis boulevard Arago, 75014, Paris, France, Email: seymour@iap.fr}
\author{Ian M$^{\rm c}$Hardy}
\author{Katherine Gunn}
\author{Derek Moss}
\affil{School of Physics \& Astronomy, University of Southampton, Highfield, Southampton, SO17 1BJ, UK Emails: imh@astro.soton.ac.uk, kfg@astro.soton.ac.uk, dm@astro.soton.ac.uk}

\begin{document}


\maketitle

\begin{abstract}
The up-turn in Euclidean normalised source counts below 1\,mJy at 1.4\,GHz 
is well established in many deep radio surveys. There are 
strong reasons, observationally and theoretically, to believe that this 
up-turn is due to strong evolution of the starforming population up to z=2.
However this hypothesis needs further confirmation spectroscopically and 
the examples in the literature are sparse. Theoretically the up-turn is 
well modelled by the evolution of the local radio
starforming population and is consistent with the up-turn seen in recent 
mid-infrared source counts at $15\,\mu$m ({\it ISOCAM}) and $24\,\mu$m 
({\it Spitzer}) and the tight correlation of the radio and MIR Luminosities 
of starforming galaxies.
\end{abstract}

\section{Introduction}

Early radio surveys were dominated by very distant, and hence very 
luminous, sources which could only be powered by very energetic 
processes. It is now accepted that this process is accretion of 
in-falling material onto a super-massive blackhole (and often accompanied 
by large-scale radio jets). As sensitivities 
improved many {\it normal} galaxies were detected in the radio. Here 
the radio emission is thought to be due from starformation 
\citep[ie synchrotron radiation and HII regions,][]{normalgalaxies}.

\section{20cm Faint Source Counts}

\begin{figure}
\centering
%\vspace{4cm}
\includegraphics[width=0.7\linewidth,angle=270]{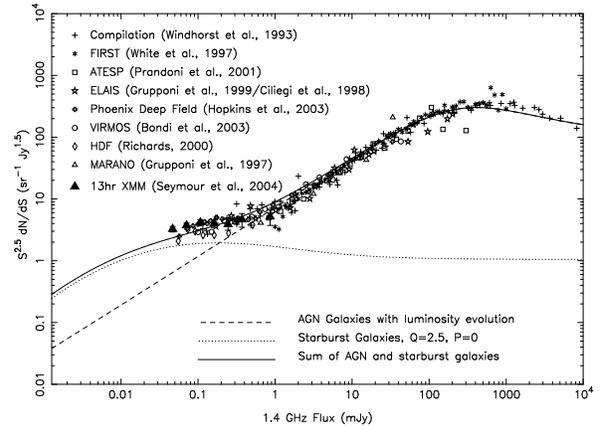}
\caption{20cm source counts from the literature showing the up-turn 
at sub-mJy flux densities and the models of two populations: AGNs 
and starforming galaxies. \label{fig:counts}}
\end{figure}

Euclidean normalised source counts (ie $S^{2.5}\times$dN/dS)  have for a 
long time shown an up-turn below 1mJy \citep{windhorst90}. This is now well 
characterised by many surveys (see Fig~\ref{fig:counts}). Current models 
\citep{king04,seymour04} explain this as the emergence of a rapidly involving 
starforming population. This idea is supported by the rapid rise in the 
starformation rate density \citep[as derived from observations at many 
wavelengths, eg][]{hopkins04}. Additionally, the similar rapid 
evolution of source counts in the MIR \citep{appleton04,pozzi04} and the 
well-known correlation of the IR 
and radio luminosities of starforming galaxies suggests that  
starformation is the major radio emission process of the faintest radio 
sources. However the contribution of AGN in the radio below 1mJy is not 
well known as a) there are few direct observations of un-ambiguous AGN 
at faint flux densities and b) the reason why only $10\%$ of AGN are strong 
radio emitters is not well understood \citep[although there are now some more 
sophisticated models around, eg ][]{jarvis04}. The simple picture of two 
unrelated 
populations (AGNs and starforming galaxies) is most likely not true, 
especially if AGN activity is triggered by mergers which also induce 
starformation. The exact contribution to the source counts from AGN 
and starforming galaxies can only be determined from direct observations 
of the individual radio sources over the full SED range.

\section{The 13hr ROSAT/{\it XMM} Survey}
\vspace{-0.1cm}

Our survey \citep{seymour04} is one of many (eg GOODS, COSMOS etc.) with 
deep radio 
observations complimented by deep observations at many other wavelengths. 
Crucially these all have deep X-ray, optical \& IR data which 
is vital in determining the true nature of the radio emission. Very 
preliminary analysis of the 13hr ROSAT/{\it XMM} Survey indicates 
that the faint radio population may, indeed, be due to starformation. 
This hypothesis is suggested by the increase of {\it radio-loudness} 
($S_{20cm}/S_{R}$) with increasing redshift and increasing luminosity for
the non-AGN population. This result can be explained by the fact that 
more luminous starbursts, more common in the past, suffer from higher 
optical extinction due to more dust. AGN show a similar trend, but are 
clearly offset from the starforming population.

The identifications of starforming galaxies will be essential in 
determining the evolution of this population and their contribution 
to the evolution of the starformation density rate which is of particular 
importance as radio is an unbiased tracer of starformation (in the 
absence of an AGN contribution). Another possibility is that very distant, 
steep radio spectrum AGN maybe contribute slightly to the faint radio 
source population (eg sources with $L_{20cm}\sim10^{24}$, AGNs, can be 
detected out to $z\sim7$ in our sample).

\vspace{-0.1cm}
\section{Evolution of Starforming Radio Luminosity Function}
\vspace{-0.1cm}

As an example of what can begin to be done we present a sample of faint radio 
sources with optical spectra ($S_{20cm}>30\mu Jy$ \& $R<21.8$) 
where we have weeded out strong AGN (ie by using the optical spectra, X-ray 
luminosity/spectra or radio morphology). We can 
then use this sample to derive an {\it evolving} starforming radio 
luminosity function which we can compare to various models. We continue 
with the caveat that a few sources may be contaminated by weak AGN. A more  
robust separation of AGN activity and starformation will be aided when the 
Spitzer data is available and we have a full SED. Spitzer is sensitive 
enough that it is likely to detect 
all the radio sources, even those without optical counterparts.

\begin{figure}
\centering
%\vspace{4cm}
%\includegraphics[width=0.8\linewidth]{lf_new.ps}
\includegraphics[width=0.7\linewidth,angle=270]{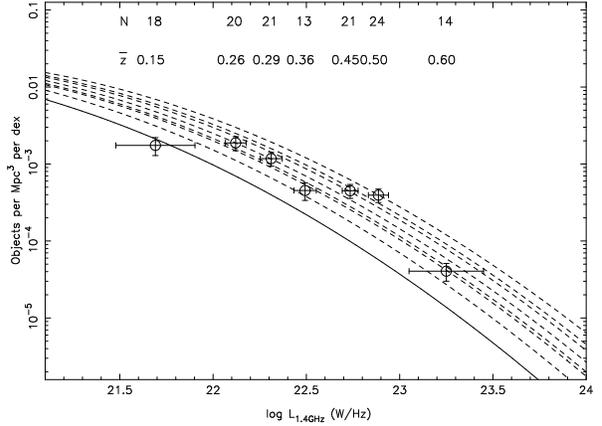}
\caption{Evolving starforming luminosity function as derived from our 
current data. The numbers at the top of the plot indicate the number 
of sources and the mean redshift of each bin. The local luminosity 
function (solid line) and successive luminosity functions 
at the redshifts indicated (dashed line) are plotted using the 
best fit evolution parameters.\label{fig:lf}}
\end{figure}
\vspace{-0.1cm}

The luminosity function is derived using the standard $1/V_{max}$ method 
where $V_{max}$ is the {\it accessible} volume determined by the two 
detection limits and is presented in Fig~\ref{fig:lf}. The number of 
objects in each bin and their mean redshift are indicated for each point. 
Objects in successively more luminous bins lie at successively higher 
redshifts. The solid shows the local starforming radio luminosity function 
\citep{sadler02} 
and the dotted lines represent this luminosity function at the redshift 
of each bin with luminosity evolution of the form $L(z)\propto(1+z)^{2.7}$ 
and density evolution of $\rho(z)\propto(1+z)^{0.15}$ \citep{hopkins04}. 

\vspace{-0.1cm}
\section{Discussion}
\vspace{-0.1cm}

The evolution of the luminosity function in Fig~\ref{fig:lf} is well fitted 
by the evolution presented (despite the clear incompleteness of the sample 
at high and low luminosities). However many different models and different 
combinations of density and luminosity evolution could fit the data presented 
here. Hence we are in the process of determining redshifts (spectroscopic 
and photometric) of optically fainter radio sources to create a larger sample. 
Using this sample we will be able to break the degeneracy of different 
models and the relative importance of 
luminosity and density evolution by accurate determination of the non-local 
luminosity function, the source counts and the starformation density rate.

%\section*{Acknowledgments}

% The following bibliography was produced with
%   \bibliographystyle{aa}
%   \bibliography{mnemonic,bibliography}

\vspace{-0.3cm}

\begin{thebibliography}{}

\bibitem[{{Appleton} {et~al.}(2004){Appleton}, {Fadda}, {Marleau}, {Frayer},
  {Helou}, {Condon}, {Choi}, {Yan}, {Lacy}, {Wilson}, {Armus}, {Chapman},
  {Fang}, {Heinrichson}, {Im}, {Jannuzi}, {Storrie-Lombardi}, {Shupe},
  {Soifer}, {Squires}, \& {Teplitz}}]{appleton04}
{Appleton}, P.~N., {Fadda}, D.~T., {Marleau}, F.~R., {et~al.} 2004, Astrophys. J., Suppl. Ser., 154,
  147

\bibitem[{{Condon}(1992)}]{normalgalaxies}
{Condon}, J.~J. 1992, ARAA., 30, 575

\bibitem[{{Hopkins}(2004)}]{hopkins04}
{Hopkins}, A. 2004, AJL., in press, astroph/0407170

\bibitem[{{Jarvis} \& {Rawlings}(2004)}]{jarvis04}
{Jarvis}, M.~J. \& {Rawlings}, S. 2004, in "Science with the Square Kilometer
  Array", eds. C. Carilli and S. Rawlings, New Astronomy Reviews (Elsevier:
  Amsterdam)

\bibitem[{{King} \& {Rowan-Robinson}(2004)}]{king04}
{King}, A.~J. \& {Rowan-Robinson}, M. 2004, MNRAS, 349, 1353

\bibitem[{{Pozzi} {et~al.}(2004){Pozzi}, {Gruppioni}, {Oliver}, {Matute}, {La
  Franca}, {Lari}, {Zamorani}, {Serjeant}, {Franceschini}, \&
  {Rowan-Robinson}}]{pozzi04}
{Pozzi}, F., {Gruppioni}, C., {Oliver}, S., {et~al.} 2004, AJ, 609, 122

\bibitem[{{Sadler} {et~al.}(2002){Sadler}, {Jackson}, {Cannon}, {McIntyre},
  {Murphy}, {Bland-Hawthorn}, {Bridges}, {Cole}, \& {Colless}}]{sadler02}
{Sadler}, E.~M., {Jackson}, C.~A., {Cannon}, R.~D., {et~al.} 2002, MNRAS, 329,
  227

\bibitem[{{Seymour} {et~al.}(2004){Seymour}, {McHardy}, \& {Gunn}}]{seymour04}
{Seymour}, N., {McHardy}, I.~M., \& {Gunn}, K.~F. 2004, MNRAS, 352, 131

\bibitem[{{Windhorst} {et~al.}(1990){Windhorst}, {Mathis}, \&
  {Neuschaefer}}]{windhorst90}
{Windhorst}, R., {Mathis}, D., \& {Neuschaefer}, L. 1990, in ASP Conf. Ser. 10:
  Evolution of the Universe of Galaxies, 389--403


%\bibitem[Allen(1973)]{allen73}
%Allen C., 1973, Astrophysical Quantities, Athlone Press

%\bibitem[Nobody et~al.(1997)]{nobody97}
%Nobody B., Somebody G., Who M.E., et~al., 1997, ApJ 331, 902

%\bibitem[Smith \& Jones(1996)]{smith96}
%Smith A., Jones B., 1996, A\&A 555, 999

\end{thebibliography}
% The results are inserted directly here to simplify
% the demonstration.

\end{document}